\documentclass[conference,8pt]{IEEEtran}
\newtheorem{th1}{Theorem}
\newtheorem{th2}{Definition}
\newtheorem{corollary}{Corollary}

\ifCLASSINFOpdf
\usepackage[pdftex]{graphicx}
\graphicspath{{../pdf/}{../jpeg/}}
\DeclareGraphicsExtensions{.pdf,.jpeg,.png}
\else
\usepackage[dvips]{graphicx}
 \graphicspath{{../eps/}}
 \DeclareGraphicsExtensions{.eps}
\fi
\usepackage[cmex10]{amsmath}

\usepackage{mdwmath}
\usepackage{mdwtab}

\IEEEoverridecommandlockouts
\begin{document}
%
\title{Secret Key Agreement Using Conferencing in State- Dependent Multiple Access Channels with An Eavesdropper
}
\author{\IEEEauthorblockN{Mohsen Bahrami, Ali Bereyhi, Mahtab Mirmohseni and Mohammad Reza Aref \thanks{This work was partially supported by Iranian NSF under contract no. $88114/46-2010$.}}
\IEEEauthorblockA{Information Systems and Security Lab (ISSL),\\
Sharif University of Technology, Tehran, Iran,\\
Email: \{bahramy, bereyhi\}@ee.sharif.edu, m.mirmohseni@ece.ut.ac.ir, aref@sharif.edu }
}


%


\maketitle

\begin{abstract}
In this paper, the problem of secret key agreement in  state-dependent multiple access channels with an eavesdropper is studied. For this model, the channel state information is non-causally available at the transmitters; furthermore, a legitimate receiver observes a degraded version of the channel state information. The transmitters can partially cooperate with each other using a conferencing link with a limited rate. In addition, a backward public channel is assumed between the terminals. The problem of secret key sharing consists of two rounds. In the first round, the transmitters wish to share a common key with the legitimate receiver. Lower and upper bounds on the common key capacity are established. In a special case, the capacity of the common key is obtained. In the second round, the legitimate receiver agrees on two independent private keys with the corresponding transmitters using the public channel. Inner and outer bounds on the private key capacity region are characterized. In a special case, the inner bound coincides with the outer bound. We provide some examples to illustrate our results. \\
\end{abstract}


\section{Introduction}
Secure communication in a network is possible when legitimate users have access to some secret keys. In [1], Shannon demonstrated that the perfect secrecy condition can be satisfied if:
\begin{align}
H(K) \geq H(M),  \nonumber
\end{align}
where, $H(K)$ and $H(M)$ are the entropies of the message and the key, respectively. Secret key generation in a network requires the existence of common randomness between users. A simple model for common randomness, in the information theory context, is distributed correlated sources. This model was first studied by Ahlswede and Csiszar  [2], where legitimate users utilize two correlated sources as common randomness to share a secret key in a noiseless network that must be concealed from an eavesdropper. In [3], new bounds on the secret key capacity over a multiterminal network with public channel were established by Gohari and Anantharam. In their model, there are $M$ legitimate terminals and an eavesdropper that have access to correlated sources. Only some of the legitimate terminals can transmit over the channel. All the legitimate terminals intend to agree on a common key that must be kept secret from the eavesdropper. In a noisy channel, common randomness can be obtained by implementing the channel distribution. This model is useful when illegal users have no access to the common randomness or a part of it. But if the legitimate users do not have any advantages compared to the illegal users, this common randomness is not beneficial for secret key sharing any more. Maurer solved this problem using a backward public channel in the wiretap model [4]. The backward public channel is a noiseless channel used by the receivers to transmit messages to the transmitters where the messages can be observed by an eavesdropper. In [2], Ahlswede and Csiszar showed that a forward noiseless channel does not help to solve the problem. In addition to the Maurer's solution, the problem can be solved when correlated sources are distributed between legitimate users in a noisy network. This idea was recently developed by Khisti $ et \ al. $ in the wiretap channel where the transmitter and the legitimate receiver have access to correlated sources [5]. Salimi and Skoglund, in another recent work, investigated the problem of secret key agreement over generalized multiple access channels using correlated sources [6]. In this channel, each of transmitters intends to agree on a private key with a receiver. Furthermore, when a forward public channel is available, the secret key sharing problem was studied by Salimi $ et \ al. $ [7]. In their model, the transmitters intend to share private keys over the generalized multiple access channel with the receiver using the public channel. The authors established examples to show that the forward public channel can improve the secret key capacity. In addition, they showed that using the forward public channel for key sharing is more effective than compress and forward strategy which was proposed in [8]. In state-dependent noisy networks, the Channel State Information (CSI) can be used as common randomness when illegal users have limited access to the CSI. In these networks, the CSI may be available causally or non-causally at the legitimate users. Khisti $et \ al.$ studied the  problem of secret key agreement over 2-receiver broadcast channels with causal or non-causal CSI where the transmitter upon observing the CSI generates a secret key and sends the required information over the channel and the legitimate receiver estimates the secret key  [9]-[11]. 

Cooperation can be effective for common key sharing in a network where there are more than one transmitter. Conferencing is one of the schemes that can be utilized to provide cooperation. In most cases, a noiseless channel with a limited rate is used to establish the conferencing scheme. In [12], Willems used the conferencing scheme in a multiple access channel where there is an interactive noiseless channel with a limited rate between the transmitters. Upon receiving sequences from the noiseless channel, each transmitter determines the channel input as a function of its message and the observed sequences.
\vspace{-1mm}
\subsection*{Main Contributions and Organization}{
Consider a multiterminal network with $n+1$ users, where one of them acts as a Trusted Center (TC) and others act as End Nodes (ENs). In addition, there is an illegal user in the network which wishes to eavesdrop. In this network, the ENs try to establish a confidential connection with the TC. Therefore, they first need to agree on some keys with the TC to announce themselves as trusted users. For transmitting the confidential message, the TC needs to generate $n$ independent private keys and share them with each of the ENs. These private keys provide an ability of multiplexing in the network. The eavesdropper tries to find the keys and attack the network. Motivated by the above scenario, we define our system model. As Fig. 1 illustrates, we consider a three-user network with an eavesdropper, in which two ENs and a TC are modeled as two transmitters and a legitimate receiver, respectively. The transmitters and the legitimate receiver are connected by a State-Dependent Multiple Access Channel (SD-MAC) where a conferencing link is available between the transmitters. In addition, the eavesdropper observes the channel. An \emph{insecure} backward public channel with an unlimited capacity is available between all the terminals. In order to achieve a secure connection; at first, the transmitters intend to share a common key over the SD-MAC with the legitimate receiver using the conferencing scheme. Then, the legitimate receiver shares an independent private key with each of the transmitters over the public channel.\\

In this model, we investigate the problem of secret key agreement in two rounds. In the first round, we establish the lower and upper bounds on the common key capacity. The intuition behind the lower bound comes from the superposition coding and random binning. The state is utilized to generate the common key by means of the hybrid joint source channel coding. In the second round, the inner and outer bounds are derived on the private key capacity region. The double random binning is used to satisfy the secrecy constrains. In this round, the private key capacity is obtained for some special cases. Different systems can be modeled as the SD-MAC with an eavesdropper. For example, we consider a binary memory with stuck at faults in which two end nodes utilize this memory to share a common key with a trusted center where an eavesdropper has access to the memory. As another example, we discuss the key agreement in the modulo-additive SD-MAC with an eavesdropper.\\

The rest of the paper is organized as follows. In Section II, the problem definition is described. In Section III, our main results and the intuitions behind them are given. In Section IV, examples are provided. Finally, proofs are presented in Section~V. 

\section{Problem Definition}{
Throughout the paper, we denote a discrete random variable with an upper case letter (e.g., $X$) and its realization by the lower case letter (e.g., $x$). We denote the probability density function of $X$ over $\mathcal{X}$ with $p(x)$ and the conditional probability density function of $Y$ given $X$ by $p(y|x)$. Finally, we use $Y^n$ to indicate vector $(Y_{1},Y_{2}, \ldots ,Y_{n})$.

A discrete memoryless SD-MAC with an eavesdropper is defined by a channel input alphabet $\mathcal{X}_1 \times \mathcal{X}_2$, a channel state alphabet $\mathcal{S}$, a channel output alphabet $\mathcal{Y}$, an eavesdropper's output alphabet $\mathcal{Z}$ and a transition probability function $p(y,z|x_1,x_2,s)$ where $\mathcal{X}_1 , \mathcal{X}_2 , \mathcal{Y} , \mathcal{Z} \ \text{and} \ \mathcal{S}$ are finite sets. As Fig.~1 illustrates, the transmitters have access to the exact CSI while the legitimate receiver has access to the degraded version of the CSI non-causally. \\

\begin{figure}
\includegraphics[width=3.6in,keepaspectratio]{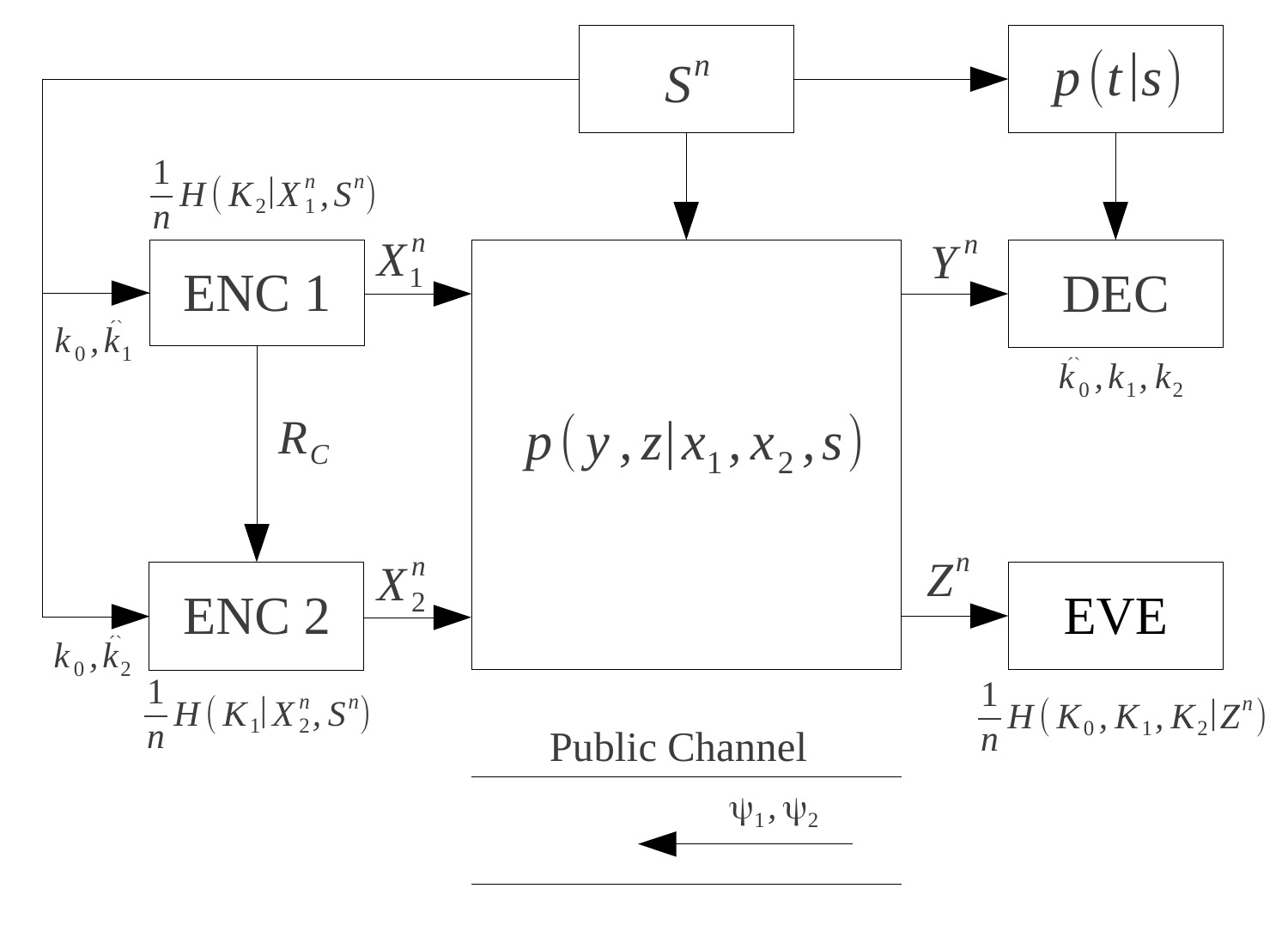}
\caption{The state-dependent multiple access channel with an eavesdropper.}
\end{figure}

We consider the interactive key agreement in the SD-MAC with an eavesdropper where a backward public channel with unlimited capacity is available from the receivers to the transmitters. We assume that a noiseless channel, with limited rate $R_C$, is available between the encoders which can be used for conferencing. The interactive key agreement scheme consists of two rounds. In the first round, the transmitters generate a common key using conferencing and transmit required information for common key sharing to the legitimate receiver via the SD-MAC. In the second round, the legitimate receiver agrees on a private key with each transmitter using the public channel. In the following we clarify the schemes with details.
\subsection{The First Round}{
In the first round, as Fig.~1 illustrates, the first transmitter, upon observing $s^n$, generates ${k_{01} \in {[1:2^{nR_0})}}$ as a common key and sends the required information to the second transmitter over the noiseless channel with limited rate $R_C$. The second transmitter generates ${k_{02} \in {[1:2^{nR_0})}}$, as a function of received information and $s^n$, to share with the legitimate receiver. Then, the transmitters determine $x_{1i}$ and $x_{2i}$ for ${i \in [1:n]}$, as deterministic functions of the corresponding common keys and $s^n,$ and transmit $x^n_1$ and $x^n_2$ over the SD-MAC with an eavesdropper. The legitimate receiver observes the channel output $y^n$ and reconstructs the common key $\hat{k}_{0}$. The sequence $z^n$ is received from the SD-MAC by the eavesdropper. 

\begin{th2}
In the first round, a rate $R_0$ is said to be achievable if for every $\epsilon > 0$ and sufficiently large $n$, there exists a protocol such that
\begin{align}
&\Pr \{ K_{01} \neq K_{02} \} < \epsilon \label{1} \\
&\Pr \{\cup_{i=1}^2 \{ \hat{K}_{0} \neq K_{0i} \} \} < \epsilon \label{2}\\
&\frac{1}{n} I(K_0;Z^n) < \epsilon \label{3} \\
&\frac{1}{n} \log \mid K_0 \mid < \frac{1}{n} H(K_0) + \epsilon \label{4} \\
&\frac{1}{n} H(K_0) > R_0 - \epsilon \label{5}
\end{align}
\end{th2}
Equation \eqref{1} investigates conferencing achievement. Equation \eqref{2} is the reliability condition of the common key. Equation \eqref{3} implies that the eavesdropper has effectively no information about the common key. Finally, the set of equations \eqref{4} and \eqref{5} investigate the uniformity conditions.
\begin{th2}
The common key capacity is the set of all achievable rates $R_0$.
\end{th2}
}
\subsection{The Second Round}{
In the second round, as Fig.~1 illustrates, the legitimate receiver, upon observing $Y^n$ and $T^n$, determines two independent private keys ${k_{1} \in [1:2^{nR_1})}$ and ${k_{2} \in [1:2^{nR_2})}$ for sharing with the first and second transmitter, respectively. The legitimate receiver transmits ${\psi_{1}=\psi_{1}(k_1)}$ and ${\psi_{2}=\psi_{2}(k_2)}$ over the backward public channel. For ${i = 1,2}$, the $i$th transmitter estimates its private key $\hat{k}_i$. The eavesdropper utilizes $z^n, \ \psi_{1} \ \text{and} \ \psi_{2}$ for eavesdropping.

\begin{th2}
In the second round, a rate pair $(R_1,R_2)$ is an achievable private key rate pairs if for every $ \epsilon >0$ and sufficiently large n there exists a protocol such that
\begin{align}
&\Pr \{ K_{i} \neq \hat{K}_{i} \} < \epsilon  \label{6} \\
&\frac{1}{n} I(K_i;Z^n,\psi_{1},\psi_{2}) < \epsilon \label{7} \\
&\frac{1}{n} I(K_i;X^n_{i^c},K_{i^c},S^n,\psi_{1},\psi_{2}) < \epsilon  \label{10}
\end{align}
\begin{align}
&\frac{1}{n} \log \mid K_i \mid < \frac{1}{n} H(K_i) + \epsilon  \label{8} \\
&\frac{1}{n} H(K_i) > R_i - \epsilon \label{9}
\end{align}
\end{th2}for $i=1,2$, where $i^c$ is $i \text{'s}$ complement, i.e., $\{ i^c , i \} = \{1,2 \}$. Equation \eqref{6} is the reliability conditions. Equations \eqref{7} and \eqref{10} mean that the eavesdropper and each transmitter have efficiently no information about the other transmitter's private key. Finally, the set of equations \eqref{8} and \eqref{9}, investigate the uniformity conditions. 

\begin{th2}
The private key capacity region is the set of all achievable rate pairs $(R_1,R_2)$.
\end{th2}
}
}
\section{Main Results}{
Here, we provide inner and outer bounds on the secret key capacity region of the SD-MAC with an eavesdropper, in two sub-sections. In sub-section \ref{sec:sec31}, we discuss the lower and upper bounds on the common key capacity. The inner and the outer bounds on the private capacity region are given in sub-section \ref{sec:sec32}. 

\subsection{The First Round}\label{sec:sec31}{
In this sub-section, we present two theorems. Theorem 1 states a lower bound on the common key capacity.

\begin{th1}[Common Key Lower Bound]
The common key rate $R_0$ is achievable for the first round if 
\begin{align}
&R_0 \leq [I(V;Y,T|U)-I(V;Z|U)]^+
\end{align}
subject to the constraints:
\begin{align}
&I(U;Y|T) \leq I(U;S) \\
&I(V;Y,T|U) \leq I(V;S|U) \\
&R_C  \geq H(U,V|S)
\end{align}
for some input distribution:
\begin{align}
&p(s,t,u,v,x_1,x_2,y,z)=p(s)p(t|s)p(u|s)p(v|u,s) \nonumber \\
&p(x_1|u,v,s)p(x_2|u,v,s)p(y,z|x_1,x_2,s),
\end{align}
\end{th1}where ${[x]^+=\max \{ x,0\}}$.

\begin{IEEEproof}[Outline of the Proof]The achievability follows by specifying the sequence $U^n$ as a description of $S^n$. $V^n$ is generated over $U^n$ using the superposition coding. The $U^n$ and $V^n$ are shared between the transmitters by utilizing the conferencing link. The random binning is applied to satisfy the secrecy constrains. Upon observing $T^n$ and $Y^n$, the legitimate receiver estimates the common key by means of joint typicality decoding. The proof is provided in section \ref{sec:sec51}. 
\end{IEEEproof}Theorem 2 states an upper bound on the common key capacity.

\begin{th1}[Common Key Upper Bound]
For the common key sharing, any rate $R_0$ must satisfy
\begin{align}
&R_0 \leq I(X_1,X_2,S;Y,T|Z)
\end{align}
\end{th1}
\emph{Proof}: See Section \ref{sec:sec52}.

In the following, we establish the common key capacity for a special case.

\begin{corollary}
If the random variables $U, V, Y \ \text{and} \ Z$ form the Markov chain, $(U,V)$ $\rightarrow$ $Y$ $\rightarrow$ $Z$, i.e., the illegal output $Z$ is the degraded version of $Y$, the common key capacity reduces to:
\begin{align}
&R_0 \leq I(V;Y,T|U,Z)
\end{align}
subject to the constraints:
\begin{align}
&I(U;Y|T) \leq I(U;S) \\  
&I(V;Y,T|U) \leq I(V;S|U) \\ 
&R_C  \geq H(U,V|S)
\end{align}
\end{corollary}
\emph{Proof}: See section \ref{sec:sec53}.

\subsection{The Second Round}\label{sec:sec32}{
Now, the bounds on the private key capacity region are given. Theorem 3 states an inner bound on the private key capacity region. 

\begin{th1}[Private Key Inner Bound]
The private key rate pair $(R_1,R_2)$ is achievable for the second round if 
\begin{align}
R_1 \leq [\min \{&I(T_1;X_1,S|T)- I(T_1;X_2,S|T), \nonumber \\
&I(T_1;X_1,S|T)-I(T_1;Z) \}]^+ \\
R_2 \leq [\min \{&I(T_2;X_2,S|T)- I(T_2;X_1,S|T) , \nonumber \\ 
&I(T_2;X_2,S|T)-I(T_2;Z) \}]^+ 
\end{align}
subject to the constraints:
\begin{align}
&I(T_1;X_1,S|T) \leq I(T_1;Y|T) \\
&I(T_2;X_2,S|T) \leq I(T_2;Y|T)
\end{align}
for some input distribution:
\begin{align}
&p(s,t,t_1,t_2,x_1,x_2,y,z)=p(s)p(x_1,x_2|s) \nonumber \\
&p(y,z|x_1,x_2,s)p(t)p(t_1|t)p(t_2|t).
\end{align}
\end{th1}
\begin{IEEEproof}[Outline of the Proof] In order to achieve the inner bound, two conditionally independent sequences $T^n_1$ and $T^n_2$ are generated with probability distribution $p(t_1|t)p(t_2|t)$. Then, we use the double random binning to satisfy the secrecy constrains. The proof is given in section \ref{sec:sec54}.
\end{IEEEproof} Theorem 4 states an outer bound on the private key capacity region.

\begin{th1}[Private Key Outer Bound]
For the private key sharing, any rate pair $(R_1,R_2)$ must satisfy:
\begin{align}
&R_1\leq \min \{I(T_1;X_1,S|Z),I(T_1;X_1|X_2,S)\} \\
&R_2\leq \min \{I(T_2;X_2,S|Z),I(T_2;X_2|X_1,S)\} 
\end{align}
\end{th1}
This bound can be directly deduced from Theorem 1 in [2].
In the following, we obtain the private key capacity region for a special case.
\begin{corollary}
If the inputs and output of the SD-MAC with an eavesdropper form a Markov chain as $(X_1,T_1)$ $\rightarrow$ $(S,T)$ $\rightarrow$ $(X_2,T_2)$ $\rightarrow$ $Z$, the private key capacity region reduces to:
\begin{align}
& R_1 \leq I(T_1;X_1|S,X_2) \\
& R_2 \leq I(T_2;X_2|S,X_1)
\end{align}
subject to the constraints:
\begin{align}
&I(T_1;X_1,S|T) \leq I(T_1;Y|T) \\
&I(T_2;X_2,S|T) \leq I(T_2;Y|T)
\end{align}
\end{corollary}
\emph{Proof}: The achievability follows from Theorem 3 where we have:
\begin{align}
&R_1 \geq I(T_1;X_1,S|T)-\max \{ I(T_1;X_2,S|T),I(T_1;Z) \}  \nonumber \\
&\stackrel{(a)}{=} I(T_1;X_1,S|T)-I(T_1;X_2,S|T)=I(T_1;X_1|S,T) \nonumber \\
&-I(T_1;X_2|S,T)=I(T_1;X_1,X_2|S,T)\nonumber \\
&-I(T_1;X_2|X_1,S,T)-I(T_1;X_2|S,T) \nonumber \\
&\stackrel{(b)}{=} I(T_1;X_2|S,T)+I(T_1;X_1|X_2,S,T)-I(T_1;X_2|S,T)\nonumber \\
&=I(T_1;X_1|X_2,S,T) \stackrel{(c)}{=} I(T_1;X_1|X_2,S), 
\end{align}
where $(a)$, $(b)$ and $(c)$ can be deduced from the Markov chain, $(X_1,T_1)$ $\rightarrow$ $S \rightarrow T$ $\rightarrow$ $(X_2,T_2)$ $\rightarrow$ $Z$. The proof of converse can be obtained from the outer bound of Theorem~4.
}
}
\section{Examples} 
Different examples can be established to illustrate our proposed model. In this section, we present some examples to explain our results.

\subsection{Binary Memory with Stuck at Faults}
Consider a network where two ENs intend to share a common key ${k_0 \in [1:2^{nR_0})}$ with the TC. In this network, a binary memory with stuck at faults is available where the eavesdropper has access to this memory. Suppose only the ENs have access to the defect information. For the key agreement, the ENs utilize the fault pattern to share the required information with the TC. The binary memory with stuck at faults can be modeled as the state-dependent memoryless channel where each of the memory cells sticks at $0$ with a probability $\frac{p}{2}$, likewise, sticks at 1 with a probability $\frac{p}{2}$ and behaves as a noiseless binary channel with a probability $1-p$ [13]. For the described example, the following argument shows that a lower bound on the common key capacity is $R_0 \leq p$ bits subject to the constraint $H(V|S) \leq 1-p$. \\

We propose a protocol for the common key agreement: Fix distribution $p(v,s)$ such that $H(V|S) \leq 1-p$. Generate a set of $2^n$ binary sequences $v^n(m_v)$, ${m_v \in {[1:2^{n})}}$ according to a Bernoulli distribution with success probability $\frac{1}{2}$ where there are roughly $2^{n(1-p)}$ sequences $v^n$ that match any given fault pattern. Partition them into $2^{nR_0}$ equal size subsets. Choose the sequence $v^n$ such that $v^n$ and $s^n$ are jointly typical respect to $p(v,s)$. Set the subset index of chosen $v^n$ as the common key. By using the above protocol and Theorem~1, we prove $R_0 \leq p$.

\emph{Proof:} By setting $U=T=\emptyset$ in Theorem~1, we have:
\begin{align}
R_0 &\leq I(V;Y)-I(V;Z)=H(V|Z)-H(V|Y) \nonumber \\
&\leq H(V)-H(V|Y) =1-(1-p)=p \ \text{bits} \nonumber
\end{align}
and for the constraint we have:
\begin{align}
I(V;Y) \leq I(V;S) &\Rightarrow H(V|S) \leq H(V|Y) \nonumber \\ 
&\Rightarrow H(V|S) \leq 1-p \nonumber \ \text{bits}
\end{align}
In fact, the ENs utilize the fault pattern for  common key sharing with the TC. Therefore, the common key rate is bounded by error probability $p$.

\subsection{The Modulo-Additive SD-MAC}
Consider the binary SD-MAC with channel output ${Y=X_1 \oplus X_2 \oplus S \oplus N_1}$ and eavesdropper's output $Z=X_1 \\ \oplus X_2 \oplus S \oplus N_2$ where ${N_1 \sim Bern(p_1)}$, ${N_2 \sim Bern(p_2)}$, ${0 \leq p_1 \leq p_2 \leq \frac{1}{2}}$ and the channel state $S$ has a Bernoulli distribution with success probability $p_S$. In proposed model, the transmitters intend to share a common key ${k_0 \in [1:2^{nR_0})}$ with the legitimate receiver $Y$ using conferencing with limited rate $R_C$. A lower bound on the common key capacity of the modulo-additive SD-MAC with eavesdropper is 
\begin{align}
R_0 &\leq H_b((\alpha \ast p_S) \ast p_1)+H_b(p_S \ast p_2)\nonumber \\
&-H_b((\alpha \ast p_S) \ast p_2)-H_b(p_S \ast p_1) \nonumber
\end{align}
subject to the constraint:
\begin{align}
&H(V|S) \leq \min \{ H_b(\alpha)+H_b(\alpha \ast p_1)-H_b((\alpha \ast p_S) \ast p_1) , R_C \} \nonumber
\end{align}

\emph{Proof:} In order to prove the lower bound, we set $U=T= \emptyset$ and $X_1=X_2=V \sim Bern(\alpha)$, using the conferencing link, in Theorem 1 such that
\begin{align}
&R_0 \leq I(V;Y)-I(V;Z) = H(Y)+H(Z|V)-H(Z) \nonumber \\
&-H(Y|V)=H(V \oplus S \oplus N_1)+H(V \oplus S \oplus N_2|V) \nonumber \\
&-H(V \oplus S \oplus N_2)-H(V \oplus S \oplus N_2|V) \nonumber \\
&=H(V \oplus S \oplus N_1)+H(S \oplus N_2)-H(V \oplus S \oplus N_2) \nonumber \\
&-H(S \oplus N_2)=H_b((\alpha \ast p_S) \ast p_1) \nonumber \\
&+H_b(p_S \ast p_2)-H_b((\alpha \ast p_S) \ast p_2)-H_b(p_S \ast p_1) \nonumber
\end{align}
for the constraint we have:
\begin{align}
& I(V;Y) \leq I(V;S) \Rightarrow H(Y)-H(Y|V) \leq H(V)-H(V|S) \nonumber \\
& H_b((\alpha \ast p_S) \ast p_1)-H_b(\alpha \ast p_1) \leq H_b(\alpha)-H(V|S) \nonumber \\
& H(V|S) \leq H_b(\alpha)+H_b(\alpha \ast p_1)-H_b((\alpha \ast p_S) \ast p_1). \nonumber
\end{align}
and
\begin{align}
&R_C \geq H(V|S) \nonumber
\end{align}where $a \ast b = a(1-b)+(1-a)b$ and $H_b(x)=-x \log(x)-(1-x) \log(1-x)$.

\section{Proofs}{
In this section, we present proofs of the main results. In order to prove Theorem 1, we employ the superposition coding [14] and random binning [15]. The intuition behind the proof of Theorem 3 comes from the Slepian \& Wolf coding [16] and the double random binning. The proofs of the outer bounds are similar to [2]. 

\subsection{Proof of Theorem 1}\label{sec:sec51}{
Fix probability distribution $p(x_{i},u,v|s) \ \text{for} \ i=1, \ 2$,

\emph{Codebook Generation}: Randomly and independently generate $2^{n{R_U}}$ sequences ${U^n}(m_u), \ {m_u \in [1:2^{n{R_U}}})$ each according to $\prod_{i=1}^n p(u_i)$. The set of all sequences $U^n$ is represented by $\mathcal{C}^U$. For each $U^n(m_u)$, randomly and conditionally independently generate $2^{n{\tilde{R}_V}}$, sequences $V^{n}(m_v), \ m_v \in [1:2^{n{\tilde{R}_V}})$, each according to $\prod_{i=1}^n p(v_i|u_i)$ and randomly partition them into $2^{n{R_V}}$ bins. Consequently, each bin consists of $2^{n({\tilde{R}_V - R_V})}$ sequences $V^n$ in average. Codebook $\mathcal{C}^U$ contains of $2^{nR_U}$ sub-codebooks where sub-codebook $m_u$ is represented by $\mathcal{C}^V(m_u)$.

\emph{Encoding}: The first encoder, ENC 1, upon observing CSI $s^n$ chooses a pair $(m_u,m_v)$ such that,
\begin{align}
&(s^n,u^n(m_u),v^n(m_v)) \in \mathcal{T}^{(n)}_{\epsilon}(S,U,V).
\end{align}
If there is no such pair, ENC 1 sets $(m_u,m_v)=(1,1)$. If there are more than one pair, ENC 1 randomly chooses $m_u \ \text{and} \ m_v$. Then, ENC 1 sends pair $(m_u ,m_v)$ over the noiseless channel with limited rate $R_C$. ENC 2 reconstructs $u^n(m_u) \ \text{and} \  v^n(m_v)$, using the codebook. The reconstruction can be done successfully if:
\begin{align}
&R_C \geq \max_{m_u} \{ \frac{1}{n} \log ( \parallel \mathcal{A}(U) \cap C^U \parallel \times \parallel \mathcal{A}(V) \cap C^V({m_u}) \parallel) \} \label{11}
\end{align}
the sets $\mathcal{A}(U)$ and $\mathcal{A}(V)$ are defined as below:
\begin{align}
&\mathcal{A}(U)= \{ u^n | \exists \ s^n \ni (u^n,s^n) \in \mathcal{T}^{(n)}_{\epsilon}(S,U) \} \nonumber \\
&\mathcal{A}(V)= \{ v^n | \exists \ s^n,u^n \ni (v^n,u^n,s^n) \in \mathcal{T}^{(n)}_{\epsilon}(S,U,V) \}. \nonumber
\end{align}where $\parallel \mathcal{A}(U) \cap C^U \parallel$ and $\parallel \mathcal{A}(V) \cap C^V({m_u}) \parallel$ indicate the number of sequences $U^n$ in $\mathcal{C}^U$ and $V^n$ in $\mathcal{C}^V(m_u), m_u \in {[1:2^{nR_U})}$ that are jointly typical with $S^n$, respectively. It can be shown that condition \eqref{11} is satisfied by $R_C \geq H(U,V|S)$. Simply, we can consider $R_C= R_U+\tilde{R}_V$.

ENC $j$ transmits $x_{ji}=x_{ji}(s_i,u_i(m_u),v_i(m_v)) \ \text{for} \ i \in {[1:n]}, \ j=1,2$ over the SD-MAC. This can be done with an arbitrarily small probability of error as $n \to \infty$ if:
\begin{align}
&R_U \geq I(U;S) \\
&\tilde{R}_V \geq I(V;S|U)
\end{align}
The above conditions can be deduced from the covering lemma [17].

\emph{Remark 1}: According to the channel distribution, the output distribution $p(y)$ can be written as
\begin{align}
&p(y)= \sum_{x_1,x_2,s} p(y|x_1,x_2)p(x_1,x_2,s). \nonumber
\end{align}
In order to cover the probability space of variables $X_1$, $X_2 $ and $S$ completely, we must generate the codewords $X^n_{1}$ and $X^n_{2}$ as functions of $S^n$.

\emph{Common Key Generation}: The transmitters choose the bin index of $ v^n(m_v)$ as the common key to share with the legitimate receiver.

\emph{Decoding}: The legitimate decoder, upon observing the channel output $ y^n$, estimates $u^n(\hat{m}_u)$ such that
\begin{align}
&(y^n,t^n,u^n(\hat{m}_u)) \in \mathcal{T}^{(n)}_{\epsilon}(Y,T,U)
\end{align}
and recovers $v^n(\hat{m}_v) \in \mathcal{C}^V(\hat{m}_u)$ such that
\begin{align}
&(y^n,t^n,u^n(\hat{m}_u),v^n(\hat{m}_v)) \in \mathcal{T}^{(n)}_{\epsilon}(Y,T,U,V).
\end{align}
If an error occurs, the legitimate decoder sets $(\hat{m}_u,\hat{m}_v)=(1,1)$. By using the packing lemma and mutual packing lemma [18], the probability of error tends to zero as $n \to \infty$ if:
\begin{align}
&R_U \leq I(U;Y|T) \\ 
&\tilde{R}_V \leq I(V;Y,T|U)  \\
&R_U + \tilde{R}_V \leq I(U,V;Y|T) 
\end{align}

\emph{Secrecy Analysis}: In order to check the security condition on the common key rate averaged over the random codebook assignments $\mathcal{C}^V$, we have:
\begin{align}
&I(K_0;Z^n|\mathcal{C}^V) \leq I(K_0;Z^n,U^n|\mathcal{C}^V)=H(K_0|\mathcal{C}^V) \nonumber \\
&-H(K_0|Z^n,U^n,\mathcal{C}^V)= H(K_0|\mathcal{C}^V)-H(K_0,V^n|Z^n,U^n,\mathcal{C}^V) \nonumber \\
&+H(V^n|K_0,Z^n,U^n,\mathcal{C}^V)=nR_V-H(V^n|Z^n,U^n,\mathcal{C}^V)\nonumber \\
&-H(K_0|Z^n,U^n,V^n,\mathcal{C}^V)+H(V^n|K_0,Z^n,U^n,\mathcal{C}^V) \nonumber \\
&\stackrel{(a)}{=} nR_V-H(V^n|Z^n,U^n,\mathcal{C}^V)+H(V^n|K_0,Z^n,U^n,\mathcal{C}^V) \nonumber \\
&\stackrel{(b)}{\leq}  nR_V-n\tilde{R}_V+nI(V;Z|U)+n\tilde{R}_V-nR_V\nonumber \\
&-nI(V;Z|U)+n\epsilon= n\epsilon
\end{align}where $(a)$ follows from the fact that $K_0$ is the bin index of $V^n$ and the equality $H(K_0|Z^n,U^n,V^n,\mathcal{C}^V)=0$ holds. $(b)$ can be deduced from inequalities $H(V^n|Z^n,U^n,\mathcal{C}^V) \leq n\tilde{R}_V-nI(V;Z|U)+n\epsilon$ and $H(V^n|K_0,Z^n,U^n,\mathcal{C}^V) \leq n\tilde{R}_V-nR_V-nI(V;Z|U)+n\epsilon$ if $R_0 \leq I(V;Y,T|U)-I(V;Z|U)+\epsilon$, the proof is similar to [18, Lemma 22.3].

\subsection{Proof of Theorem 2}\label{sec:sec52}{
In our described model, the legitimate receiver should be able to estimate the common key $K_0$ correctly, therefore, according to the Fano's inequality we have $\frac{1}{n}H(K_0|Y^n,T^n) \leq \epsilon$ and also, the security condition $\frac{1}{n}I(K_0;Z^n) \leq \epsilon$ must be satisfied.
We obtain an upper bound on $R_0$,
\begin{align}
&nR_0=H(K_0)= I(K_0;Z^n)+H(K_0|Z^n) \nonumber \\ 
&\stackrel{(a)}{\leq} H(K_0|Z^n)+n\epsilon=I(K_0;Y^n,T^n|Z^n) \nonumber \\
&+H(K_0|Y^n,T^n,Z^n)+n\epsilon \stackrel{(b)}{\leq}  I(K_0;Y^n,T^n|Z^n)+2n\epsilon \nonumber \\
&\leq I(K_0,X^n_1,X^n_2,S^n;Y^n,T^n|Z^n)+2n\epsilon \nonumber \\
&\stackrel{(c)}{=} I(X^n_1,X^n_2,S^n;Y^n,T^n|Z^n)+2n\epsilon \nonumber \\ 
&\leq \sum_{i=1}^nI(X_{1i},X_{2i},S_i;Y_i,T_i|Z_i)+2n\epsilon \nonumber \\
&\stackrel{(d)}{=} nI(X_{1Q},X_{2Q},S_Q;Y_Q,T_Q|Z_Q,Q)+2n\epsilon \nonumber \\
&=nI(X_{1Q},X_{2Q},S_Q;Y_Q,T_Q|Z_Q)+2n\epsilon
\end{align}
where $(a)$ deduces from the security condition, $(b)$ follows from the Fano's inequality, $(c)$ can be derived from the Markov chain $K_0 \rightarrow (X_1^n,X_2^n) \rightarrow (Y^n,Z^n)$. $(d)$ can be obtained by considering $Q$ as a uniform variable over ${[1:n]}$.
}
\subsection{Proof of Corollary 1}\label{sec:sec53}{
The achievability follows from Theorem~1 where we have:
\begin{align}
&R_0 \geq I(V;Y,T|U)-I(V;Z|U)=I(V;Y,T,Z|U) \nonumber \\
&-I(V;Z|U,Y,T)-I(V;Z|U)\stackrel{(a)}{=}I(V;Z|U) \nonumber \\
&+I(V;Y,T|Z,U)-I(V;Z|U)=I(V;Y,T|Z,U)
\end{align}
where $(a)$ comes from the Markovity. For the converse proof, we have:
\begin{align}
&nR_0=H(K_0)=I(K_0;Z^n)+H(K_0|Z^n) \nonumber \\
& \stackrel{(a)}{\leq} H(K_0|Z^n)+n\epsilon = I(K_0;Y^n,T^n|Z^n) \nonumber \\
&+H(K_0|Y^n,T^n,Z^n)+n\epsilon \stackrel{(b)}{\leq} I(K_0;Y^n,T^n|Z^n)+2n\epsilon \nonumber \\
& \leq I(K_0,X^n_1,X^n_2,S^n;Y^n,T^n|Z^n)+2n\epsilon \nonumber \\
& \stackrel{(c)}{\leq} \sum_{i=1}^n I(K_0,X^n_1,X^n_2,S^n;Y_i,T_i|Z^i,Y^n_{i+1},T^n_{i+1})+2n\epsilon \nonumber \\
& \stackrel{(d)}{=} \sum_{i=1}^n I(K_0,X^n_1,X^n_2,S_i;Y_i,T_i|Z_i,Y^n_{i+1},T^n_{i+1},S^{i-1}) \nonumber \\
&+2n\epsilon \stackrel{(e)}{=} \sum_{i=1}^n I(V_i;Y_i,T_i|Z_i,U_i)+2n\epsilon \nonumber \\
&\stackrel{(f)}{=}nI(V_Q;Y_Q,T_Q|Z_Q,U_Q,Q)+2n\epsilon \nonumber \\
&=nI(V_Q;Y_Q,T_Q|Z_Q,U_Q)+2n\epsilon
\end{align}
where $(a)$ follows from the security condition $I(K_0;Z^n) \leq n\epsilon$, $(b)$ comes from the Fano's inequality $H(K_0|Y^n,T^n) \leq n\epsilon$. $(c)$ and $(d)$ can be deduced from the Markov chains, $(K_0,X^n_1,X^n_2,S^n)$ $\rightarrow$ $(T^n_{i+1},Y^n_{i+1})$ $\rightarrow$ $Z^n_{i+1}$ and $(K_0,X^n_1,X^n_2)$ $\rightarrow$ $S^{i-1}$ $\rightarrow$ $Z^{i-1}$, respectively. By defining $V_i=(K_0,X^n_1,X^n_2,S_i)$ and $U_i=(Y^n_{i+1},T^n_{i+1},S^{i-1})$, $(e)$ is obtained. $(f)$ can be established by considering $Q$ as a uniform variable over $[1:n]$.
\subsection{Proof of Theorem 3}\label{sec:sec54}{
Fix probability distribution $p(t_1,t_2,t)= p(t)p(t_1|t)p(t_2|t)$ 

\emph{Codebook Generation}: For $i=1,2$, randomly generate $2^{nR_{T_i}}$ sequences $T^n_{i}(m_{t_i}), \ m_{t_i} \in {[1:2^{nR_{T_i}})}$ each according to $\prod_{j=1}^n p(t_{ij}|t)$ and partition them into $2^{nR'_{T_i}}$ bins and $2^{nR''_{T_i}}$ sub-bins using the double random binning. Therefore, there are $2^{n(R_{T_i}-R'_{T_i})}$ sequences $T^n_{i}$ in each bin and $2^{n(R_{T_i}-R'_{T_i}-R''_{T_i})}$ sequences $T^n_{i}$ in each sub-bin in average. $\mathcal{C}^{T_i}$ indicates the coodebook containing all $T^n_i$. The bin $m'_{t_i}$ and sub-bin $m''_{t_i}$ are represented by $\mathcal{B}'(m'_{t_i}) \ \text{and} \ \mathcal{B}''(m''_{t_i})$, respectively.

\emph{Private Key Generation}: For $i=1, 2$, the legitimate receiver, $Y$, upon observing the channel output $y^n$, chooses $t^n_i(m_{t_i})$ such that
\begin{align}
&(t^n_{i}(m_{t_i}), y^n) \in \mathcal{T}^{(n)}_\epsilon (T_i,Y)
\end{align}
and sets $m''_{t_i}, \ t^n_{i}(m_{t_i}) \in \mathcal{B}''(m''_{t_i})$, as private key $k_i$ to share with the $i\text{th}$ transmitter. 
This can be done with an arbitrarily small probability of error if:
\begin{align}
&R_{T_i} \geq I(T_i;Y|T).
\end{align}
The above conditions can be deduced directly from the covering lemma.

\emph{Use of Public Channel}: The legitimate receiver, $Y$, transmits $m'_{t_i}, \ t^n_{i}(m_{t_i}) \in \mathcal{B}'(m'_{t_i}) \ \text{for} \ i=1,2$ over the backward public channel.

\emph{Key Reconstruction}: The $i\text{th}$ transmitter, upon receiving $m'_{t_i}$, estimates $\hat{m}_{t_i}$ such that
\begin{align}
&t^n_{i}(\hat{m}_{t_i}) \in \mathcal{B}'(m'_{t_i}), \\
&(x^n_{i},s^n,t^n_{i}(\hat{m}_{t_i})) \in \mathcal{T}^{(n)}_\epsilon (X_i,S,T_i),
\end{align}
and finds $k_i$ such that $t^n_{i}(\hat{m}_{t_i}) \in \mathcal{B}''(k_i)$.\\
This can be done with an arbitrarily small probability of error if:
\begin{align}
&R_{T_i}-R'_{T_i} \leq I(X_i,S;T_i|T) \qquad \text{for} \ i=1,2.
\end{align}
The above conditions can be deduced directly from the packing lemma.

\emph{Secrecy Analysis}: 
In order to check the security condition on the private key rate $R_1$ averaged over the random codebook assignments $\mathcal{C}^{T_1}$, we have:
\begin{align}
&I(K_1;Z^n , \psi_1 , \psi_2|\mathcal{C}^{T_1})= H(K_1|\mathcal{C}^{T_1})-H(K_1|Z^n,\psi_1 , \psi_2,\mathcal{C}^{T_1}) \nonumber \\
&= H(K_1|\mathcal{C}^{T_1})-H(K_1,T^n_1|Z^n,\psi_1 , \psi_2,\mathcal{C}^{T_1})\nonumber \\
&+H(T^n_1|K_1,Z^n,\psi_1 , \psi_2,\mathcal{C}^{T_1})=nR''_{T_1}\nonumber \\ 
&-H(T^n_1|Z^n,\psi_1,\psi_2,\mathcal{C}^{T_1}) -H(K_1|Z^n,T^n_1,\psi_1,\psi_2,\mathcal{C}^{T_1}) \nonumber \\
&+H(T^n_1|K_1,Z^n,\psi_1 , \psi_2,\mathcal{C}^{T_1}) \stackrel{(a)}{=} nR''_{T_1}\nonumber \\
&+H(T^n_1|K_1,Z^n,\psi_1 , \psi_2,\mathcal{C}^{T_1})- H(T^n_1|Z^n,\psi_1,\psi_2,\mathcal{C}^{T_1}) \nonumber \\
&\stackrel{(b)}{\leq} nR''_{T_1}-nR_{T_1}+nR'_{T_1}+nI(T_1;Z)\nonumber \\
&+nR_{T_1}-nR'_{T_1}-nR''_{T_1}-nI(T_1;Z)+n\epsilon= n\epsilon
\end{align}where $(a)$ follows from the fact that $K_1$ is the sub-bin index of $T^n_1$ and the equality ${H(K_1|Z^n,T^n_1,\psi_1,\psi_2,\mathcal{C}^{T_1})=0}$ holds. $(b)$ can be obtained from the inequalities $H(T^n_1|Z^n,\psi_1,\psi_2,\mathcal{C}^{T_1}) \leq nR_{T_1}-nR'_{T_1}-nI(T_1;Z)+n\epsilon$ and $H(T^n_1|K_1,Z^n,\psi_1 , \psi_2,\mathcal{C}^{T_1}) \leq nR_{T_1}-nR'_{T_1}-nR''_{T_1}-nI(T_1;Z)+n\epsilon$ if $R_1 \leq I(T_1;X_1,S|T)-I(T_1;Z)+\epsilon$, the proof is similar to [18, Lemma 22.3]. And, 
\begin{align}
&I(K_1;K_2,X^n_2,S^n,\psi_1,\psi_2|\mathcal{C}^{T_1}) \nonumber \\
&\leq I(K_1;K_2,X^n_2,S^n,T^n,\psi_1,\psi_2|\mathcal{C}^{T_1}) \nonumber \\
&= H(K_1|\mathcal{C}^{T_1})-H(K_1|K_2,X^n_2,S^n,T^n,\psi_1 , \psi_2,\mathcal{C}^{T_1})\nonumber 
\end{align}
\begin{align}
&= H(K_1|\mathcal{C}^{T_1})-H(K_1,T^n_1|K_2,X^n_2,S^n,T^n,\psi_1 , \psi_2,\mathcal{C}^{T_1})\nonumber \\
&+H(T^n_1|K_1,K_2,X^n_2,S^n,T^n,\psi_1,\psi_2,\mathcal{C}^{T_1}) \nonumber \\
&=nR''_{T_1}-H(T^n_1|K_2,X^n_2,S^n,T^n,\psi_1,\psi_2,\mathcal{C}^{T_1})\nonumber \\
&-H(K_1|K_2,X^n_2,S^n,T^n,T^n_1,\psi_1,\psi_2,\mathcal{C}^{T_1})\nonumber \\
&+H(T^n_1|K_1,K_2,X^n_2,S^n,T^n,\psi_1 , \psi_2,\mathcal{C}^{T_1}) \nonumber \\
&=^{(a)} nR''_{T_1}- H(T^n_1|K_2,X^n_2,S^n,T^n,\psi_1,\psi_2,\mathcal{C}^{T_1})\nonumber \\
&+H(T^n_1|K_1,K_2,X^n_2,S^n,T^n,\psi_1 , \psi_2,\mathcal{C}^{T_1})\nonumber \\
&\leq^{(b)} nR''_{T_1}-nR_{T_1}+nR'_{T_1}+nI(T_1;X_2,S|T)\nonumber \\
&+nR_{T_1}-nR'_{T_1}-nR''_{T_1}-nI(T_1;X_2,S|T)+n\epsilon= n\epsilon
\end{align}where $(a)$ follows from the fact that $K_1$ is the sub-bin index of $T^n_1$ and the equality $H(K_1|K_2,X^n_2,S^n,T^n,T^n_1,\psi_1,\psi_2,\mathcal{C}^{T_1}) \\ =0$ holds. $(b)$ can be deduced from the inequalities $H(T^n_1|K_2,X^n_2,S^n,T^n,\psi_1,\psi_2,\mathcal{C}^{T_1}) \leq nR_{T_1}-nR'_{T_1}-nI(T_1 \\ ;X_2,S|T)+n\epsilon$ and $H(T^n_1|K_1,K_2,X^n_2,S^n,T^n,\psi_1 , \psi_2,\mathcal{C}^{T_1}) \\ \leq nR_{T_1}-nR'_{T_1}-nR''_{T_1}-nI(T_1;X_2,S|T)+n\epsilon$ if $R_1 \leq I(T_1;X_1,S|T)- I(T_1;X_2,S|T)+\epsilon$, the proof is similar to [18, Lemma 22.3]. Finally, we have:
\begin{align}
R_1 \leq \min \{&I(T_1;X_1,S|T)- I(T_1;X_2,S|T), \nonumber \\
&I(T_1;X_1,S|T)-I(T_1;Z) \} \nonumber
\end{align}
Similarly, we can check the security conditions for $K_2$.
}
\section{Conclusion}{
In this paper, we investigated the problem of interactive secret key sharing over a state-dependent multiple access channel with an eavesdropper. In our proposed model, the transmitters share a common key with the receiver over multiple access channel in the first round. The conferencing scheme has a beneficial role in the common key sharing. In the second round, the receiver agrees on two independent private keys with the corresponding transmitters using the public channel. The inner and outer bounds on the capacity region have been established for the common and the private keys capacity region.
}

\begin{thebibliography}{99}{
\bibitem{rf}{C. E. Shannon, `` Communication theory of secrecy systems," \emph{Bell System Technical Journal} vol.~ 28, pp. 656-715, 1949.
\bibitem{rf} R. Ahlswede and I. Csiszár,``Common randomness in information theory and cryptography Part I: Secret sharing,” \emph{IEEE Trans.~Inf.~Theory}, vol.~39, no.~4, pp.~1121-1132, Jul. 1993.
\bibitem{rf} A. A. Gohari and V. Anantharam, ``Information-theoretic key agreement of multiple-terminals Part I: Source model," \emph{IEEE Trans.~Inf.~Theory}, vol.~56, no.~8, pp.~3973-3996, Aug. 2010.
\bibitem{rf} U. M. Maurer, ``Secret key agreement by public discussion from common information," \emph{IEEE Trans.~Inf.~Theory}, vol.~39, no.~3, pp.~733-742, May 1993.
\bibitem{rf} A. Khisti, S. Diggavi, and G. Wornell, ``Secret-key generation using correlated sources and channels," \emph{IEEE~Trans.~Inf.~Theory}, vol.~58, no.~2, pp.~652-670, Feb. 2012.
\bibitem{rf} S. Salimi and M. Skoglund, ``Secret key agreement using correlated sources over the generalized multiple access channel," \emph{Arxiv preprint}, arXiv:~1204.2922v1, Apr. 2012.  
\bibitem{rf} S. Salimi, M. Salmasizadeh, M. R. Aref, and Jovan Dj Golic, ``Key agreement over multiple access channel," \emph{IEEE Trans.~on Information Forensics and Security}, vol.~6, Issue~3, pp.~775-790, Sep.~2011.
\bibitem{rf} E. Ekrem and S. Ulukus, ``Effects of cooperation on the secrecy of multiple access channels with generalized feedback," in \emph{Proc.~42nd Ann.~Conf. Information Sciences and Systems (CISS)}, Princeton, NJ, pp.~791-796, Mar. 2008.
\bibitem{rf} A. Khisti, ``Secret key agreement on wiretap channel with transmitter side information," in \emph{Proc.~Eur.~Wireless}, Lucca, Italy, pp.~802-809, Apr. 2010.
\bibitem{rf} A. Khisti, S. Diggavi, and G. Wornell, ``Secret key agreement with channel state information at the transmitter," \emph{IEEE Trans.~on Information Forensics and Security}, vol.~6, no.~3, pp.~672-681, Sep. 2011.
\bibitem{rf}  A. Khisti, S. Diggavi, and G. Wornell, ``Secret key agreement using asymmetry in channel state knowledge," in \emph{Proc.~Int.~Symp.~Inf.~Theory}, Seoul, Korea, pp.~2286-2290, Jun.-Jul. 2009.
\bibitem{rf} F. Willems, ``The discrete memoryless multiple channel with 
partially cooperating encoders," \emph{IEEE Trans.~Inf.~Theory}, vol.~29, no.~3, pp.~441-445, May. 1983.
\bibitem{rf} C. Heegard and A. El Gammal, ``On the capacity of computer memory with defects," \emph{IEEE Trans.~Inf.~Theory}, vol.~29, no.~5, pp.~731-739, Sept. 1983. 
\bibitem{rf} T. M. Cover, ``Broadcast channels," \emph{IEEE Trans.~Inf.~Theory}, vol.~18, no.~1, pp.~2-14, Jan. 1972.
\bibitem{rf} T. M. Cover, ``A proof of the data compression theorem of Slepian and Wolf for ergodic sources," \emph{IEEE Trans.~Inf.~Theory}, vol.~21, no.~2, pp.~226-228, Mar. 1975.
\bibitem{rf} D. Slepian and J. K. Wolf, ``Noiseless coding of correlated information sources," \emph{IEEE Trans.~Inf.~Theory}, vol.~19, no.~4, pp.~471-480, Jul. 1973.
\bibitem{rf} T. M. Cover and J. A. Thomas, \emph{Elements of Information Theory}, 2nd ed.~Hoboken, NJ:~Wiley, 2006.
\bibitem{rf} A. El Gamal and Y. H. Kim, \emph{Network Information Theory}, 1st ed.~Cambridge University Press, 2011.
}
}
\end{thebibliography}

\end{document}